\documentclass{rmaa}

\title{The Low Excitation Planetary Nebulae HuDo\,1 and HuBi\,1 and their [WC10] Central Stars\altaffilmark{1}}

\author{M. Pe\~na
\affil{Instituto de Astronom\'{\i}a \linebreak
Universidad Nacional Aut\'onoma de M\'exico}}

\altaffiltext{1}{Based on data collected at the Observatorio Astron\'omico Nacional,
 San Pedro M\'artir, Ensenada, B.~C., M\'exico.}

\shortauthor{Pe\~na}
\shorttitle{The [WC\,10] PNe HuDo\,1 and HuBi\,1}

\SetVolume{41} \SetFirstPage{423} \SetYear{2005}

\ReceivedDate{2005 January 17}
\AcceptedDate{2005 May 16}

\fulladdresses{
\item Miriam Pe\~na: Instituto de Astronom\'\i{}a, UNAM,
Apdo. Postal 70-264, 04510 M\'exico D.~F., M\'exico  (miriam@astroscu.unam.mx).
}

\listofauthors{M. Pe\~na}
\indexauthor{Pe\~na,~M.}

\abstract{Low- and high-resolution spectra of the planetary nebulae HuDo\,1 and
HuBi\,1, around [WC-late] stars, are analyzed.  The objects belong to the
galactic disk, with heliocentric radial velocities of $-$12~km s$^{-1}$ (HuDo\,1) and 57~km
s$^{-1}$ (HuBi\,1).  C(H$\beta$) reddening values are of 2.04 for HuDo\,1 and 1.22 for
HuBi\,1.  Plasma line ratios are used to estimate physical conditions and
abundances.  We find $\log$(O/H)+12 = 8.43 and 8.57, and N/O = 0.2 and 0.1 for
HuDo\,1 and HuBi\,1 respectively.  HuBi\,1 is the only PN excited by a very late
[WC] star showing intense nebular \ion{He}{1} recombination lines.  From the
stellar lines we derive a [WC\,10] type for both stars, although HuBi\,1 central
star should be slightly hotter for providing a large amount of He$^0$ ionizing
photons.  Nebular and stellar parameters of HuDo\,1 and HuBi\,1 are compared
with those of other [WC\,10] objects, concluding that the stars of HuDo\,1 and
HuBi\,1 should be intrinsically fainter.  In particular HuBi\,1 seems a
low-density evolved nebula around a low-mass slowly-evolving star.  }

\addkeyword{Planetary Nebulae: individual (HuDo\,1, HuBi\,1)} 
\addkeyword{Stars: Post-AGB} 
\addkeyword{Stars:  Wolf-Rayet}

\RescaleTitleLengths{0.9}
 \begin{document}
 \maketitle

\begin{figure*}[!t]
\includegraphics[width=\textwidth]{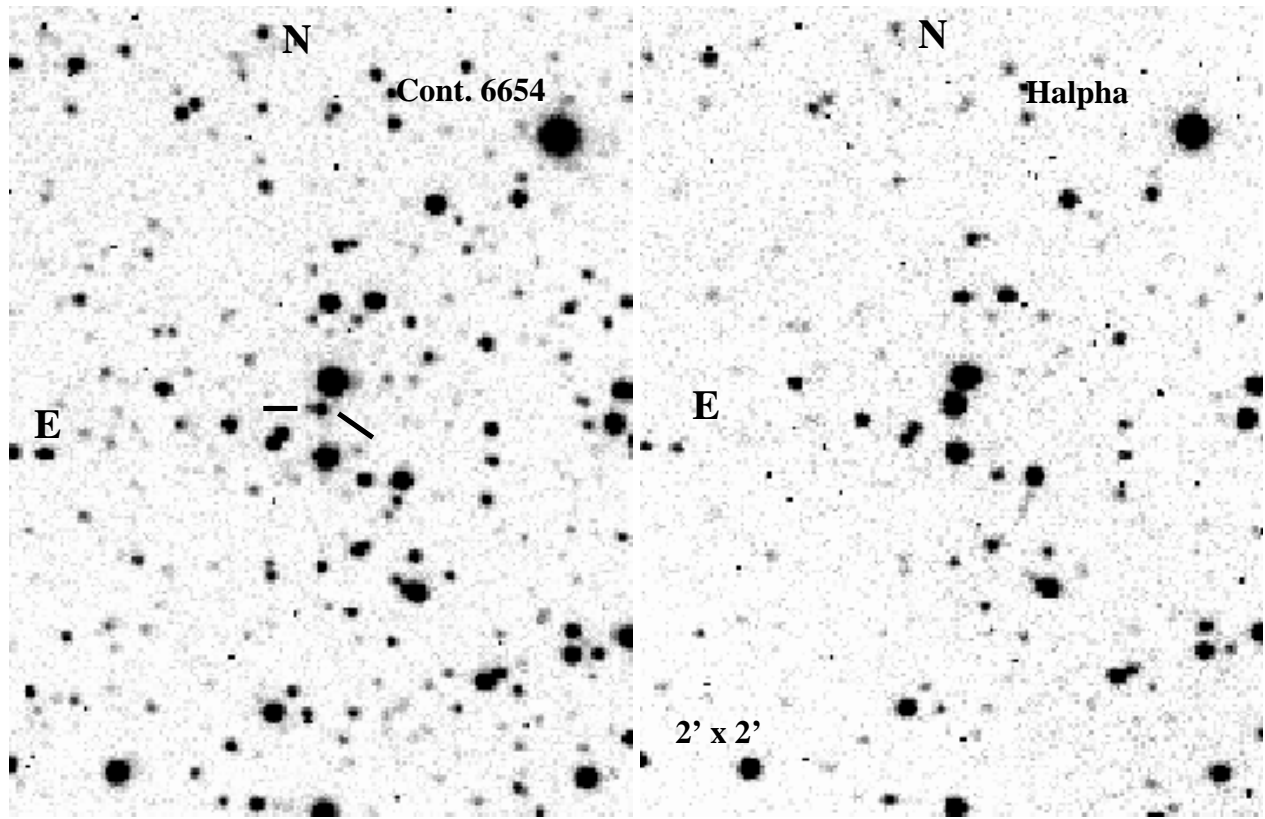}
\caption{Continuum ($\lambda$6654, 5 min exposure time) and  H$\alpha$ (10 min
 exposure time) images of a $2'\times2'$ zone around HuDo\,1. The PN is clearly detected
as a compact H$\alpha$ emitting object, with a diameter of about 1.7 arcsec.}
\end{figure*}

\begin{figure*}[!t]
\begin{center}
\begin{tabular}{ll}
\includegraphics[width=6cm,height=6cm]{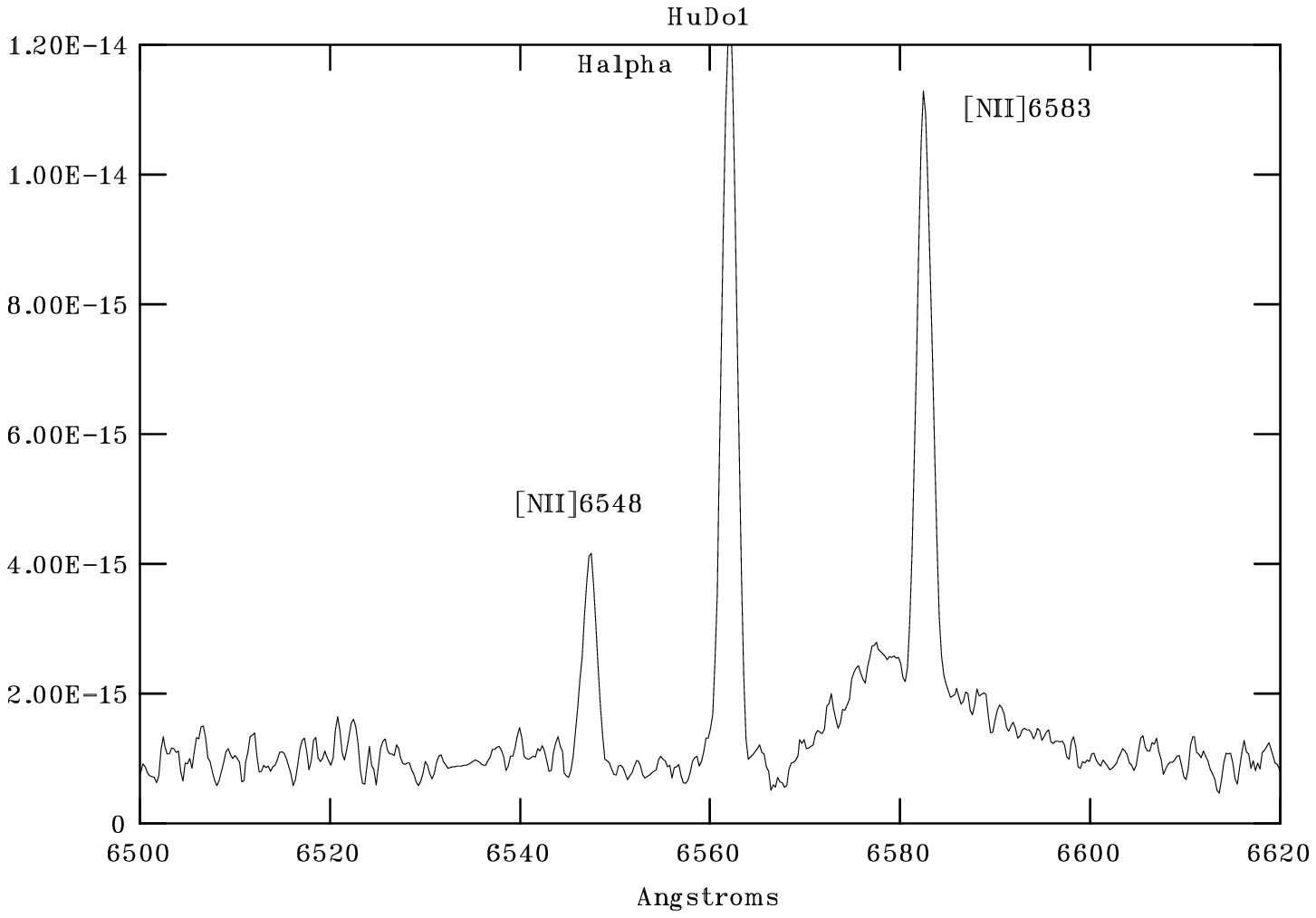}& \includegraphics[width=6cm,height=6cm]{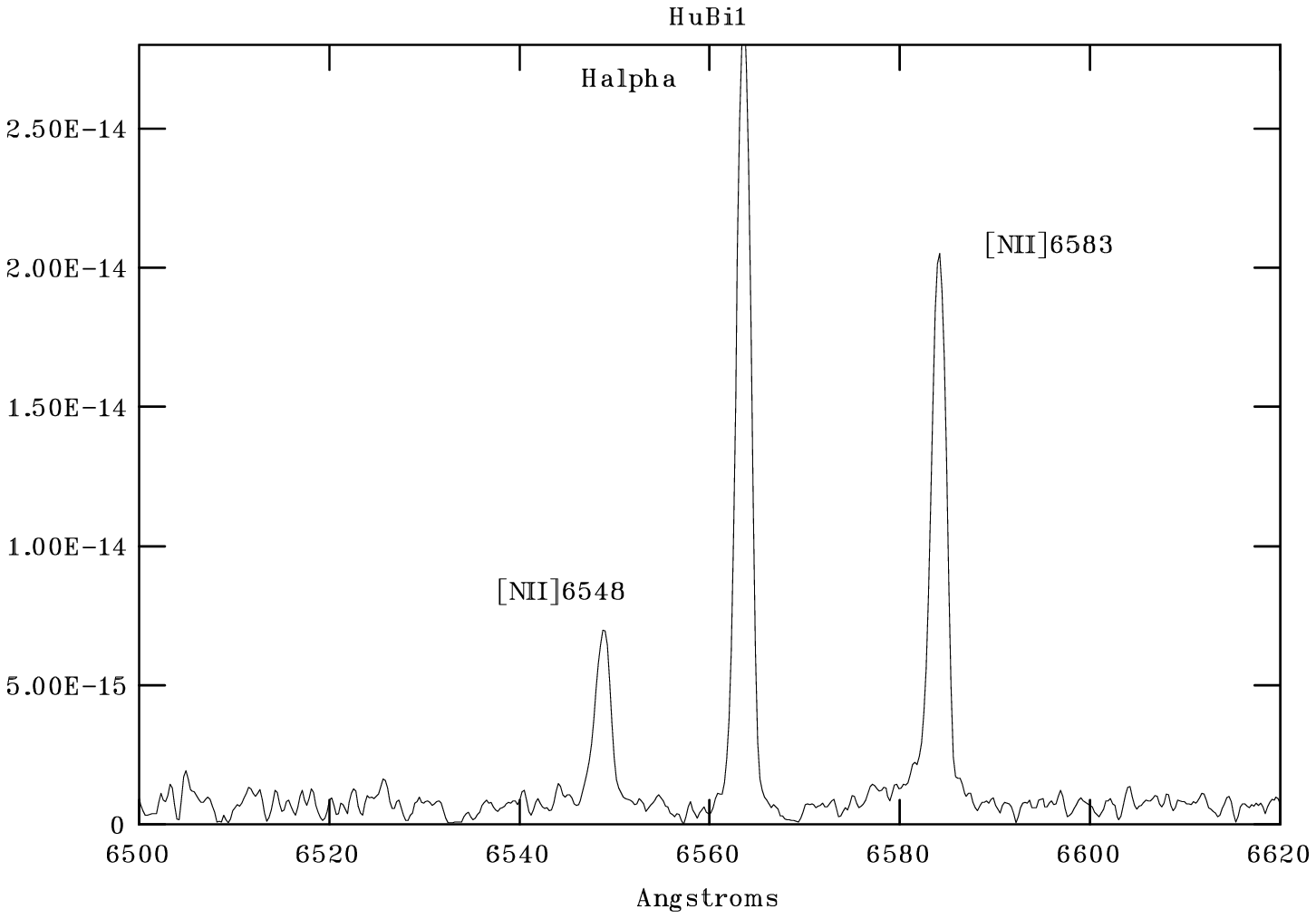}\\
& \\
\includegraphics[width=6cm,height=6cm]{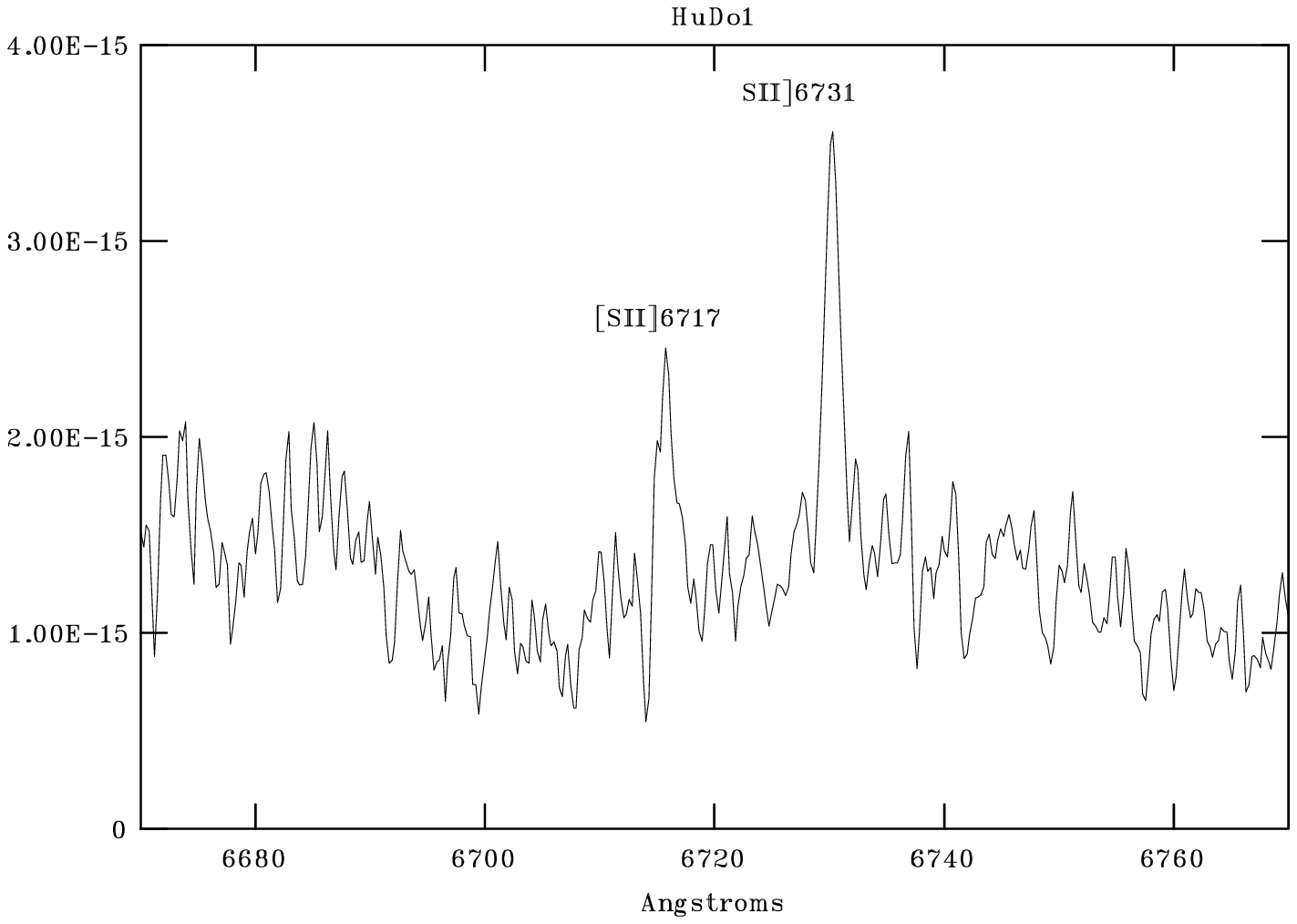}& \includegraphics[width=6cm,height=6cm]{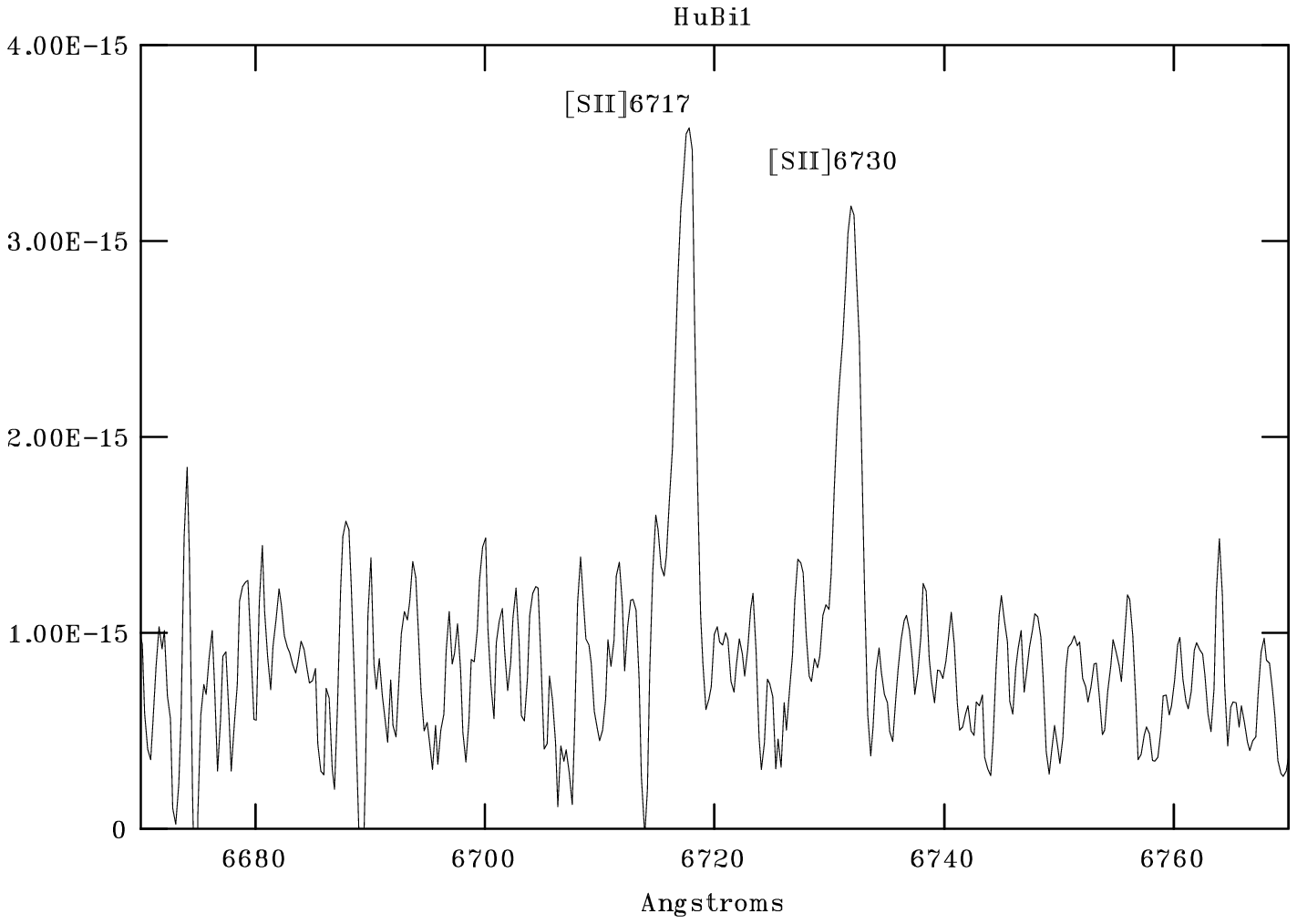}\\
& \\
\end{tabular}
\vspace{-0.5cm}
\caption{H$\alpha$ and [\ion{S}{2}] high-resolution spectral zones for HuDo\,1 and
HuBi\,1, obtained in April 2004. Fluxes are in units of erg cm$^{-2}$
s$^{-1}$. For HuDo\,1 the nebular lines are easily deblended from the wide stellar lines and
can be measured with high accuracy. The central star of HuBi\,1  is much fainter (relative to 
the nebula) and its lines are not noticeable in the high resolution spectra. }
\end{center}
\vspace{0.4cm}
\end{figure*}

\begin{figure}[!t]
\includegraphics[width=\columnwidth]{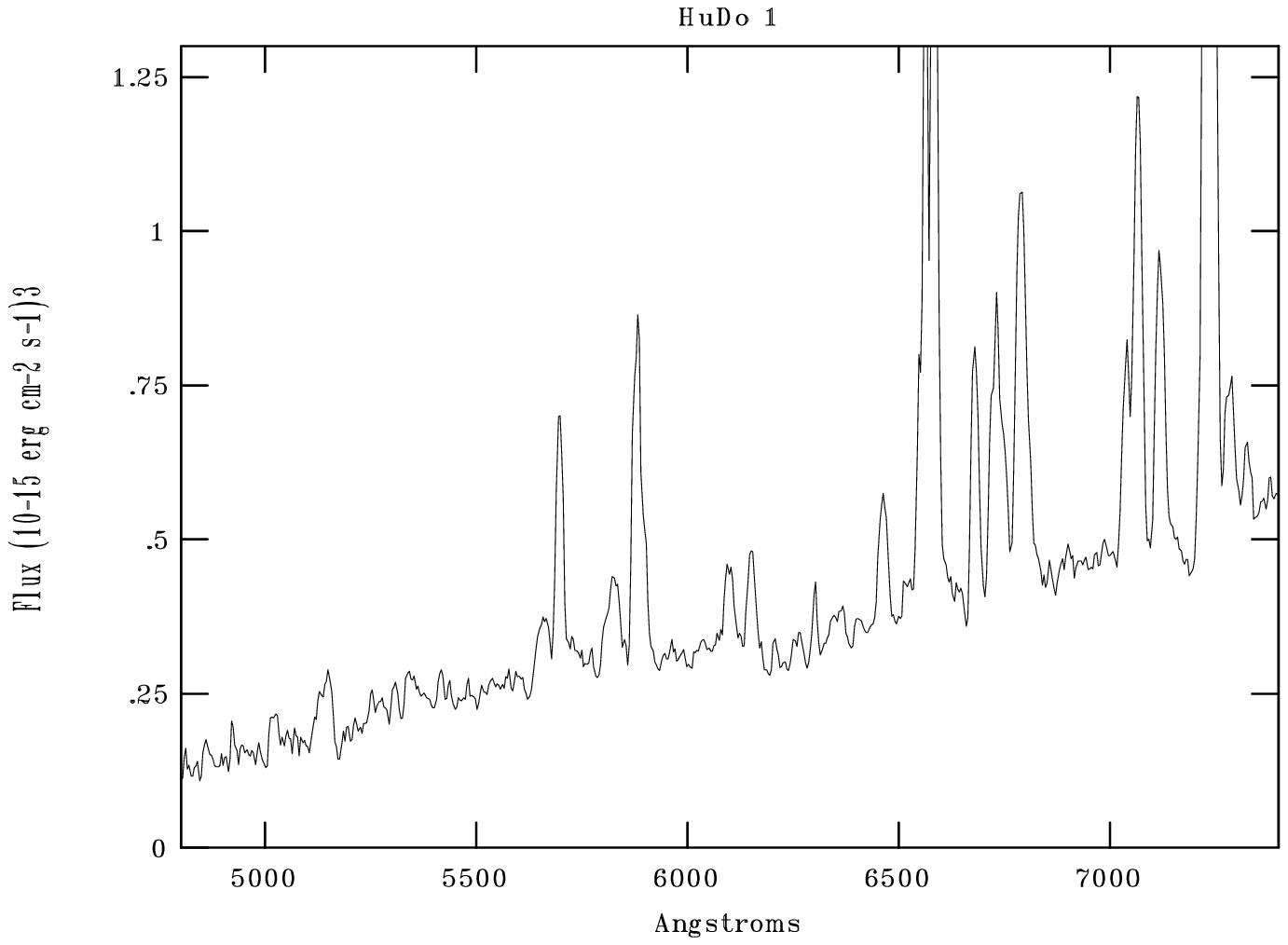}
\hfill
\includegraphics[width=\columnwidth]{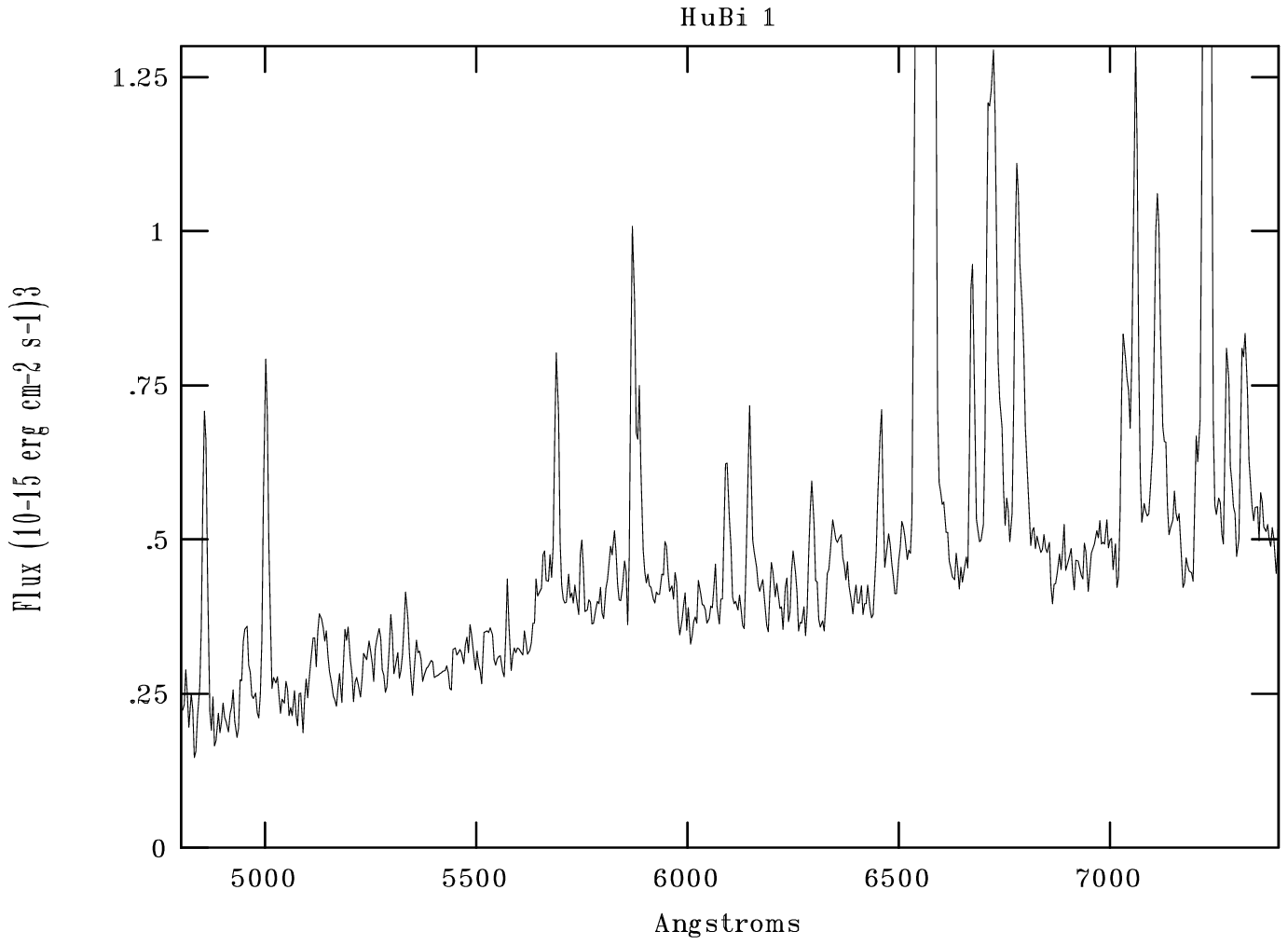}
\caption{Low-resolution spectra for HuDo\,1 and HuBi\,1, obtained in June 2002. 
Strong stellar lines are blended with the nebular emission, specially for HuDo\,1
 where the star is relatively much brighter and shows wider lines.}
\end{figure}

\begin{figure}[!t]
\includegraphics[width=\columnwidth]{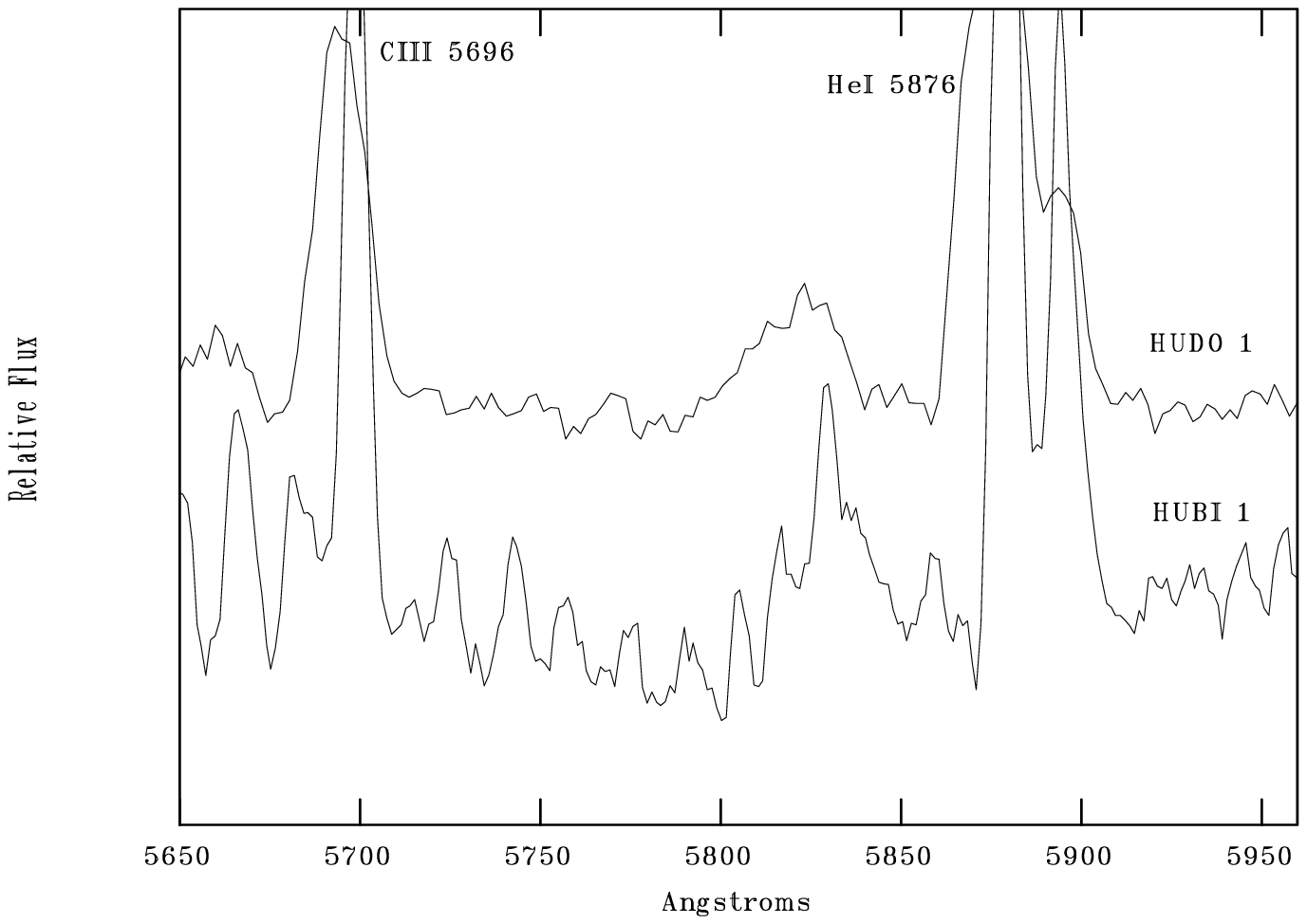}
\caption{Low-resolution stellar spectra showing the zone around \ion{C}{4}5801,5812. These lines are
in the wing of \ion{C}{3} 5826.}
\end{figure}
 
\section{Introduction}

During a spectroscopic survey of protoplanetary candidates, Hu \& Bibo (1990)
and Hu \& Dong (1992) discovered the cool [WC] nature of the central stars of
the low excitation planetary nebulae PN G012.2+04.9 (IRAS17514-1555, PM1-188,
HuBi\,1) and PN G060.4+01.5 (IRAS19367+2458, PM1-310, HuDo\,1).  Hereinafter
they will be called by their usual names:  HuBi\,1 and HuDo\,1.  Both central
stars were classified by the authors as [WC\,11] type stars, on the basis of a
qualitative analysis of the stellar spectra which are dominated by
\ion{C}{2} and \ion{C}{3} emission lines.  These very late WR type stars of the
carbon series are very infrequent.   The compilation presented by Gorny
(2001) lists only 7 planetary nebulae (PNe) with such a central star.  In addition
to HuDo\,1 and HuBi\,1, the list includes M4-18, CPD-56$^0$8031, K2-16, He\,2-113
 and Vo\,1. All these central stars were classified as [WC\,11] until the new 
quantitative classification scheme for WC and WO stars proposed by Crowther, De Marco, \& Barlow (1998)  
who moved most of these objects to the [WC\,10] type, K2-16 (PN
G352.9+11.4) remaining the only  example in their sample of a [WC\,11] star.
Regarding massive WRs, none so late WC massive star has been found so far. \\

 Another quantitative classification scheme, in principle better adapted for
[WC] central stars, was proposed more recently by Acker \& Neiner (2003).  In this
work they also move all the objects previously classified as [WC\,11] to the
[WC\,10] type, except for K2-16 and M4-18.  These authors declare that their
scheme is coherent with the one by Crowther et~al.\@ (1998).  There are however
some important differences in criteria for all the [WC] types.  Unfortunately
Acker \& Neiner criteria show discontinuities and the primary criteria are not
well defined.

In a recent paper Gorny et~al.\@  (2004) apparently have found several [WC\,11]
stars among their sample of PNe towards the galactic bulge.  However due to
their low signal-to-noise data for the central stars, Gorny et~al.
classification is based on the old qualitative scheme by Hu \& Bibo (1990).  In
addition the authors do not exclude the possibility of these stars being WELS.
The classification of these objects needs to be redone on the basis of good
quality spectra and using the new quantitative classification criteria.

The [WC]-late objects constitute the first step of the evolutionary path
proposed for the WR central stars (e.g., Gorny \& Tylenda 2000), which would
evolve from cool [WC]s ([WC\,10-11]) to hot [WC]s (presently classified as
[WO\,1-4] by Crowther et~al.\@  1998 and Acker \& Neiner 2003) to PG\,1159 white
dwarfs.  It is important therefore to understand the nature and evolutionary
status of the [WC\,10] stars.

Recent studies of the nebulae and stars for some late [WC] objects have been
performed by De Marco, Barlow, \& Storey  (1997) and De Marco \& Crowther (1999)
 (See Table~6).  For HuBi\,1, imaging and spectroscopic data were analyzed by Pollacco \& Hill
(1994) who reported a nebular diameter of about 18$''$.  According to these
authors the nebula consists of an external zone showing an electron density of
about 50 cm$^{-3}$ surrounding a higher-density central zone.  They derived N,
O, and S abundances reporting values typical of PNe, but found a possibly
enhanced He abundance.

The central star of HuBi\,1 was studied by Leuenhagen \& Hamann (1998) who, from
non-LTE expanding atmosphere models, derived $T_{\rm eff} \sim 35,000$~K,
terminal wind velocity v$_\infty$ = 360~km~s$^{-1}$, a transformed radius of 7.07
$R_\odot$ and a chemical composition (in mass percentage) of $\beta$(H) = 1,
$\beta$(He) = 42, $\beta$(C) = 50, and $\beta$(O) = 7.  By assuming a stellar
luminosity of $\log$~L $\sim$~3.70, a distance of 1.5~kpc, a mass loss rate of
$\log$~$\dot{M} \sim -5.70$, and a stellar radius $R_*$ = 1.88 $R_\odot$ were
derived for the central star by these authors.  Crowther et~al.\@  (1998) and
Acker \& Neiner (2003) analyzed the stellar lines attributing a spectral type
[WC\,10] to this star. 

In our spectroscopic program for studying systematically a large sample of PNe
around [WC] central stars (Pe\~na et~al.\@  1998;
 Pe\~na, Stasi\'nska, \& Medina 2001;
 Medina et~al.
in preparation), HuDo\,1 and HuBi\,1 were included among our targets.  High
resolution data were obtained for HuBi\,1 in 1997 and preliminary results were
published by Pe\~na et~al.\@  (2001) who reported an electron density lower than
235 cm$^{-3}$.  This density is unusually low for a nebula around such a late
[WC] star (see discussion in Pe\~na et~al.\@  2001 and in \mbox{\S 6).}

For HuDo\,1 no quantitative studies have been made apart from the qualitative
stellar classification by Hu \& Dong (1992).  We found that the ID map published
by Hu \& Dong (1992) is indicating an object which resulted to be a F8 main
sequence star, and the ID chart in the Strasbourg-ESO Catalogue of Galactic PNe
(Acker et~al.\@  1992) indicates a different but also wrong object whose spectrum
corresponds to a late-type star.  Consequently we decided to search for the
planetary nebula, using direct imaging in H$\alpha$ and nearby continuum to
detect the H$\alpha$ emitting objects in the field presented by Hu \& Dong.
With this strategy ---very useful to detect emission line objects--- we finally
found a low excitation PN whose central star, in addition, presented WC-late
type emission lines.  The right ID chart for HuDo\,1 is presented in Figure~1.

Once determined the right position of HuDo\,1, spectrophotometric data for both
objects were acquired to analyze the nebulae and central stars parameters.
Description of the observations are presented in \S 2.  In \S 3 and \S 4 we
discuss the nebular properties.  The analysis of the stellar data is presented
in \S 5, and our conclusions are discussed in \mbox{\S 6.}

\section{Observations and data analysis}

\subsection{Direct Imaging of HuDo\,1}

H$\alpha$ on-band off-band direct images for HuDo\,1 were acquired with the
2.1~m telescope of the Observatorio Astron\'omico Nacional, S.P.M., B.~C.,
M\'exico (OAN-SPM).  The CCD Site3 detector was used with interferential filters
centered at H$\alpha$ ($\lambda$6564) and continuum ($\lambda$6654) on May 27,
2001.  Exposure times were 10~min for H$\alpha$ and 5~min for $\lambda$6654.
These images allow us to identify the planetary nebula as an object different
from that indicated in the ID charts by Hu \& Dong (1992) and Acker et~al.\@
(1992).  Fig.~1 shows both images.  The planetary nebula is the weak object
marked.  It is a very compact nebula with a diameter of about 1.7~arcsec.

\setlength{\tabnotewidth}{0.8\textwidth}
\tablecols{9}
\begin{table*}[!t]\centering
\caption{$\log$ of observations for HuDo\,1 and HuBi\,1}
\begin{tabular}{lllccllll}
\toprule
Date& Spectr. & Detector$^a$ & $\Delta \lambda$ & Slit Width$^b$ &HuDo\,1$^c$ & HuBi\,1$^c$\\
\midrule
040897&ech &CCD2k & 3360--7360 & 4$''$& $\cdots$    & 2$\times$15m \\
250900&BCh &CCD2k & 4400--6750 & 4$''$& $\cdots$   & 2$\times$10m\\
260900&BCh &CCD2k & 4400--6750 & 4$''$& $\cdots$   & 2$\times$10m\\
240801&ech &CCD2k  & 3360--7360 & 4$''$& 15m, 2$\times$20m & $\cdots$ \\
260801&ech &CCD2k  & 3360--7360 & 4$''$& 2$\times$15m & $\cdots$     \\
040602&BCh &CCD-S3 & 3500--7400 & 4$''$&  3$\times$15m & 4$\times$10m \\
050602&BCh &CCD-S3 & 3500--7400 & 4$''$& 10m,15m & 2$\times$10m\\
060602&BCh &CCD-S3 & 3500--7400 & 4$''$& 3$\times$15m & 3$\times$10m \\
230404&ech &CCD-S3 & 3860--6800 & 2$''$&  $\cdots$ &   3$\times$15m    \\
240404&ech &CCD-S3 & 3860--6800 & 2$''$& 2$\times$15m& $\cdots$ \\
250404&ech &CCD-S3 & 4480--7560 & 4$''$& 4$\times$15m & $\cdots$ \\
260404&ech &CCD-S3 & 4480--7560 & 2$''$& $\cdots$ & 15m,10m \\
170804&BCh &CCD-S3 & 5600--7750& 4$''$& 3$\times$10m& $\cdots$ \\
\noalign{\smallskip}
\bottomrule
\tabnotetext{a}{CCD2k: 2048$\times$2048 pix of 14$\mu$, CCD-S3: 1024$\times$1024 pix of 24 $\mu$.
The spectral resolution is about 0.2 \AA \ with echelle and 5--6~\AA \ with B\&Ch.}
\tabnotetext{b}{Slit was always oriented E-W.}
\tabnotetext{c}{Number of exposures $\times$ exposure time.}
\end{tabular}
\vspace{0.06cm}
\end{table*}


\setlength{\tabnotewidth}{0.50\linewidth}
\tablecols{5}
\begin{table*}[!t]\centering
\tabcolsep=0.20cm
\caption{Observed and dereddened nebular line \linebreak
 fluxes, relative to H$\beta$, for HuDo\,1 and HuBi\,1}
\begin{tabular}{lr@{.}lr@{.}lr@{.}lr@{.}l}
\toprule
\noalign{\smallskip}
           & \multicolumn{4}{c}{HuDo~1} & \multicolumn{4}{c}{HuBi~1} \\
          \cline{2-5}  \cline{6-9}
 $\lambda$(\AA)  Ion&\multicolumn{2}{c}{F/H$\beta$} & \multicolumn{2}{c}{F/H$\beta$} &\multicolumn{2}{c}{F/H$\beta$} & \multicolumn{2}{c}{F/H$\beta$}   \\
             & \multicolumn{2}{c}{Observed}   & \multicolumn{2}{c}{Deredd.}    & \multicolumn{2}{c}{Observed}   & \multicolumn{2}{c}{Deredd.}     \\
\midrule
\noalign{\smallskip}
4340 H$\gamma$   &  \multicolumn{2}{c}{\nodata}    & 
\multicolumn{2}{c}{\nodata}   & 0&32 & 0&45 \\ 
5007 [\ion{O}{3}]&   $<$ 0&5 & $<$0&4 & 0&30 & 0&27 \\
5755 [\ion{N}{2}]&   $<$0&6 & $<$0&07 & 0&04:& 0&02: \\ 
5876 \ion{He}{1} &   $<$0&25 & $<$0&09  & 0&27 & 0&15 \\
6548 [\ion{N}{2}]&   3&34 & 0&73      & 1&40 & 0&57 \\
6563 H$\alpha$   &   13&52 & 2&86     & 7&12 & 2&82 \\
6583 [\ion{N}{2}]&   10&73 & 2&26     & 4&19 & 1&66 \\
6717 [\ion{S}{2}]&    1&88 & 0&38     & 0&58 & 0&22 \\
6731 [\ion{S}{2}]&    2&39 & 0&48     & 0&55 & 0&21 \\
7319 [\ion{O}{2}]&    0&89 & 0&12     & 0&29:& 0&09:\\
7330 [\ion{O}{2}]&    0&68 & 0&09     & 0&32:& 0&10:\\
\hline 
\noalign{\smallskip}
F(H$\beta$)(E-15)\tabnotemark{a}& 1&60 & 179&5    & 15&6  & 259&1 \\
EW(H$\beta$)    & \multicolumn{2}{c}{15~~~}   & \multicolumn{2}{c}{\nodata}       & 
\multicolumn{2}{l}{87}   & \multicolumn{2}{c}{\nodata} \\ 
c(H$\beta$)     & 2&05 &  \multicolumn{2}{c}{\nodata} & 1&22 & \multicolumn{2}{c}{\nodata} \\
\bottomrule
\tabnotetext{a}{ Flux in erg cm$^{-2}$ s$^{-1}$.} \\
\end{tabular}
\end{table*}

\subsection{Spectrophotometric Data}

Long-slit spectrophotometric data were gathered during different observing runs
with two different spectrographs (high and low resolution) attached to the
2.1~m telescope at OAN-SPM.  The high resolution echelle REOSC spectrograph
allows to obtain spectra with a resolution of about 20,000 ($\Delta \lambda /
\lambda$ better than 0.25) which is very useful to safely deblend the narrow
nebular lines from the wide stellar ones, specially in the case of HuDo\,1 (see
Figure~2).  The low-resolution Boller \& Chivens spectrograph was used with a
grid of 600 l/mm.  The spectral resolution for such a set-up is 2--3 \AA,
depending on the CCD pixel size.  For the 4$''$ slit used for our observations,
the spectral resolution is about 5--6 \AA.

The $\log$ of observations is presented in Table~1 where the observing date (ddmmyy),
the spectrograph (echelle or Boller \& Chivens) and the detector used are listed in
Columns  1, 2, and 3.  We present the observed spectral ranges in Column 4, in
Column 5 the slit widths, and in Columns 6 and 7, the number of exposures and the
exposure times for each object.

Data were reduced using IRAF \footnote[2]{IRAF is distributed by NOAO, which is
operated by AURA, Inc.  under contract with the National Science Fundation.}
package.  Spectra were bias and flat-field corrected, afterwards they were extracted
using an extraction window of 10 pixels, equivalent to 3.2 arcsec for the Site3
detector and 2 arcsec for the CCD2k.  Due to the small extension of HuDo\,1, the
extraction window includes the whole object.  This is not the case for HuBi\,1,
where only the central zone was extracted.  A He-Ne-Ar lamp was employed for
wavelength calibration and several standard stars from the list by Hamuy et~al.\@
(1992) for echelle spectra, and Oke (1990) for low resolution spectra, were used for
flux calibration.

Portions of the high-resolution spectra are shown in Fig.~2.  The zones around
H$\alpha$ and [\ion{S}{2}]~6717,6731 are displayed showing the narrow nebular
lines.  The wide stellar lines are only noticeable for HuDo\,1, which has a
brighter star with stronger lines.  The central star of HuBi\,1 is almost
undetectable in the echelle spectra.  Notice how the high resolution
allows to safely deblend the narrow nebular lines from the wide stellar lines
for HuDo\,1.  This is very important to obtain confident diagnostic line ratios
for plasma analysis.  Figure~3 presents calibrated low-resolution spectra for
both objects, not corrected for reddening.  In this case the stellar lines are
highly blended with the nebular ones for both objects, but specially for HuDo\,1
which presents much wider stellar lines.  Due to these objects belong to the
galactic disk, they are heavily extinguished and reddened and there is hardly
detection in the blue zone ($\lambda \leq$ 4200 \AA) of spectra.


\begin{table}[!t]\centering
\caption{Ionic and elemental abundance \linebreak ratios}
\begin{tabular}{llrr}
\toprule
\noalign{\smallskip}
 Line & Ratio             &   HuDo\,1 & HuBi\,1 \\
 \midrule
 \noalign{\smallskip}
 5876 & He$^+$/H$^+$ &  \multicolumn{1}{c}{$<$0.06}  & 0.11 \\ 
 7319 &O$^+$/H$^+$ &  2.6E-4  & 3.2E-4 \\
 7330 &O$^+$/H$^+$ &  2.5E-4  & 4.1E-4 \\
 5007 &O$^{+2}$/H$^+$ & 1.7E-5& 1.2E-5 \\
 6583 &N$^+$/H$^+$ &  5.4E-5  & 3.9E-5 \\
 6717 &S$^+$/H$^+$ &  5.3E-6 & 1.4E-6\\
 6731 &S$^+$/H$^+$ &  5.1E-6  & 1.3E-6\\
 \hline
 \noalign{\smallskip}
 \multicolumn{2}{l}{$\log$ O/H + 12} & \multicolumn{1}{r}{8.43} & 8.57 \\
 \multicolumn{2}{l}{$\log$ He/H + 12} & \multicolumn{1}{c}{$>$10.8~~} & $>$11.04\\ 
 \multicolumn{2}{l}{N/O}   & \multicolumn{1}{r}{0.21}    & 0.11  \\
 \bottomrule
 \end{tabular}
 \end{table}

\section{Radial velocities and nebular expansion}

The high-resolution spectra allow to determine the radial velocity of objects
with good accuracy.  From the observed nebular lines we have determined
heliocentric radial velocities of $-12\pm$8 km~s$^{-1}$ for HuDo\,1 and
$57\pm$8 km\,s$^{-1}$ for HuBi\,1.  These low values are indicating that both
objects are in the solar neighborhood.

The nebular lines present, in both cases, a single but well-resolved component.
Therefore the FWHM can be used as an indicator of the nebular expansion.  After
subtracting the instrumental and thermal widths (by assuming they add in
quadrature and that the electron temperature is 9400~K) the expansion velocity,
V$_{\rm exp}$, for HuDo\,1 is about 30 km~s$^{-1}$ (or lower if part of the FWHM
is caused by turbulence).  The lines of HuBi\,1 are slightly wider showing a FWHM
of 65 km~s$^{-1}$, therefore V$_{\rm exp}$ is $\leq$ 33 km~s$^{-1}$.  Pe\~na 
et~al.\@  (2001) reported a FWHM of 61 km~s$^{-1}$ for H$\beta$ in very good agreement
with the value derived here.

\section{Nebular parameters}
\subsection{Line Fluxes}
All the nebular lines detected were measured in both, low and high resolution
spectra.  High resolution data were used specially when, in the low resolution
case, the nebular lines appear blended with stellar lines.  Thus we derived
useful  line ratios for plasma diagnosis.

The observed nebular fluxes, relative to H$\beta$, are presented in Table 2.
H$\alpha$ to H$\beta$ ratios were used to derive the logarithmic reddening
correction at H$\beta$, c(H$\beta$), by assuming case B recombination theory according to
 Storey \& Hummer (1995).  The values for c(H$\beta$) are listed at the end of the
same table and were used to deredden the line flux ratios, by employing the
galactic reddening law by Seaton (1979).  The dereddened flux ratios are listed in
Columns 3 and 5 of Table~2.  These values have an uncertainty better than 20\%,
except for those cases marked with a colon, which have an uncertainty of a factor of~1.5.

Both nebulae present very similar low-excitation degree, with [\ion{O}{3}] 5007 flux
smaller than 0.5 H$\beta$ flux (this is only an upper limit for HuDo\,1, where the
line is not detected), and [\ion{N}{2}] 6583 about twice H$\beta$.  However in
HuBi\,1 the nebula is much brighter, relative to the central star, than in HuDo\,1
(see the equivalent widths of H$\beta$, EW(H$\beta$), at the end of Table~2).
Another interesting difference is the nebular \ion{He}{1} 5876 line which is
perfectly visible and presents a large intensity in HuBi\,1, while it is not
detected in HuDo\,1.  We will discuss this in the next section.

\subsection{Plasma Diagnosis: Electron Temperatures and Densities}

Plasma diagnosis was performed with the help of the task NEBULAR of IRAF.  For
HuBi\,1 the temperature sensitive line ratio [\ion{N}{2}] 5755/6583 $\sim$~0.012$\pm$0.003,
 indicates an electron temperature $T_{\rm e}~ \sim 9400 \pm
1500$~K.  The auroral line is very faint providing a large uncertainty in the
temperature. 
 [\ion{N}{2}]~5755 was not detected for HuDo\,1, and its upper
limit indicates an upper limit of 14,000 K for $T_{\rm e}$ which is certainly too
high.  We will adopt $T_{\rm e}$ = 9400~K for both nebulae.

From the density sensitive [\ion{S}{2}] 6731/6717 line ratio (obtained from high
resolution observations of April 2004), we derive for HuDo\,1 N$_{\rm e} \sim$
3300$\pm$500~cm$^{-3}$ and 800$\pm$300~cm$^{-3}$ for HuBi\,1.  This latter value
is larger than the values derived by Pe\~na et~al.\@  (2001) and Pollacco \& Hill
(1994).  This could be due to we are not measuring the same nebular zone in this
extended nebula.  However, in any case, the derived density values are unusually
low for a such a late [WC] object.


 \subsection{Chemical Abundances}

Ionic abundances for the visible ions:  O$^+$, O$^{++}$, N$^+$, S$^+$ and He$^+$
were computed by assuming electron densities of 3300 cm$^{-3}$ and 800 cm$^{-3}$
for HuDo\,1 and HuBi\,1 respectively, and $T_{\rm e}$ = 9400~K for both objects.
The results are listed in Table~3 where estimates for the O and He abundances
and the N/O ratios are also included by assuming that O/H = O$^+$/H$^+$ +
O$^{++}$/H$^+$, He/H $>$ He$^+$/H$^+$, and N/O = N$^+$/O$^+$.  The derived oxygen
abundances and the N/O abundance ratios are about solar.  The chemical abundance
pattern indicates that both objects are disk planetary nebulae.  

The He$^+$/H$^+$ abundance ratio equal to 0.11 derived for HuBi\,1 (which is
indicating a large He enhancement in this nebula), deserves a separate
discussion.  First, this is the only nebula around such a late [WC] star
presenting detectable \ion{He}{1} recombination lines (we confirmed the presence
of nebular \ion{He}{1} 5876 by extracting the nebular emission outside the star
and found a nebular spectrum similar to that presented by Pollacco \& Hill 1994
with a strong 5876 \AA; see their Figure~2b).  For none other studied PN of this
kind, the nebular \ion{He}{1} lines have been observed (for instance, De Marco
et~al.\@  1997 and De Marco \& Crowther  1999 considered that all the \ion{He}{1} 5876
emission detected in M4--18, CPD-56$^0$8031 and He2--133, is from stellar origin)
indicating that the amount of He$^+$ is very low and that most of the nebular
helium should be in neutral form.  This is expected due to the low effective
temperature (about 30 to 32~kK) determined for these central stars.  The central
star of HuBi\,1 should be hotter (Leuenhagen \& Hamann 1998 determined 35 kK)
and in some way the He$^0$ ionizing photons should not be totally consumed by
the He- and C-rich stellar atmosphere, providing enough photons to partially
ionize the nebular helium.  The grid of photo-ionization models computed by
Stasi\'nska (1982) shows that in a nebula with solar abundances and density of
10$^3$ cm$^{-3}$, a star with $T_{\rm eff} \geq $ 35,000~K (black-body
distribution) can produce a He$^+$ fraction of about 0.6, leaving 40\% in
He$^0$.  Lower stellar temperatures produces much less He$^+$ fraction, while
higher $T_{\rm eff}$ produces too large O$^{++}$ fractions.  If we adopt the
model with $T_{\rm eff}$ = 35~kK for HuBi\,1, its He/H abundance would be as large
as 0.18!  In \S 6 we discuss a compendium of nebular and stellar parameters,
including the He abundance, for all the well-studied [WC\,10] objects.

\section{The Central Stars}
\subsection{ Stellar Lines}

These lines were measured on the low resolution (Boller \& Chivens) spectra
because in the high resolution ones only the strongest stellar emission lines
were detected.  The measurements were relatively straightforward for HuBi\,1
which shows narrower stellar lines.  For HuDo\,1, in many cases the lines appear
heavily blended (see Fig.~3 and Figure~4) therefore we used the high resolution
spectra as an indication to disentangle the components.

Equivalent widths, observed and dereddened fluxes for stellar lines (relative to
\ion{C}{3} $\lambda5696$) for both objects are listed in Table~4.  As expected,
stellar spectra are dominated by \ion{C}{2} and \ion{C}{3} lines and some
\ion{He}{1} lines are also detected.  Some of the lines are blends of several
components and they have been marked with a {\it b}, immediately after the
wavelength.

From the comparison of different observations we have estimated the
uncertainties in stellar fluxes as follows:  for observed fluxes lower than 0.3
(relative to \ion{C}{3} 5696) the uncertainties are about 20--25\%; for fluxes
between 0.3 and 0.8, uncertainties are about 15\%; the uncertainties are lower
for brighter lines.  Lines marked with a colon show uncertainties of a factor of~1.5.



\begin{table*}[!t]\centering
\caption{Observed and dereddened stellar line fluxes, \linebreak
 relative to \ion{C}{3} 5696 \AA }
\begin{tabular}{lrr@{.}lr@{.}lrr@{.}lr@{.}l}
\toprule
\noalign{\smallskip}
&   \multicolumn{5}{c}{ HuDo\,1}& \multicolumn{5}{c}{HuBi\,1} \\
 \cline{2-6}  \cline{7-11}
 $\lambda$ (\AA) Ion  & \multicolumn{1}{c}{~~EW\tabnotemark{a}}
  & \multicolumn{2}{c}{F(obs)\tabnotemark{b}}&
 \multicolumn{2}{c}{F(dered)}
 &\multicolumn{1}{c}{~~EW\tabnotemark{a}}
 &\multicolumn{2}{c}{ F(obs)\tabnotemark{b}} &
\multicolumn{2}{c}{ F(dered) }\\
\midrule 
\noalign{\smallskip}
4267 \ion{C}{II}& 21 & 0&20 & 0&94 & 20 & 0&47 & 1&23 \\
4618 \ion{C}{II}&  10 & 0&1: & 0&3: & 4 & 0&13 & 0&25\\
4650\tablenotemark{b}\ion{C}{II}I& 25&0&20& 0&61& 15 &0&50 & 0&93\\
4686 \ion{He}{II}& ~~~8: & 0&1:& 0&2: & 2 & 0&1: & 0&2:\\
4922 \ion{He}{I}& 8 & 0&1:& 0&2:& 7 & 0&19&0&30\\
5015 \ion{He}{I}& 15 &0&12& 0&25 & 8 &0&21 & 0&29\\
5034\tabnotemark{b} \ion{C}{II}& 16 &0&16& 0&32& 10 & 0&33 & 0&46\\
5140\tabnotemark{b} \ion{C}{II}& 32 &0&60& 1&16& 40& 1&37 &1&76\\ 
5261  \ion{C}{II}& 15 &0&20& 0&34& 8 & 0&31 & 0&37 \\
5305 \ion{C}{III}&  8 & 0&1:&  0&16:& 4 &0&16 & 0&18 \\
5340 \ion{C}{II}&  14 &0&23& 0&35 & 8 & 0&23 & 0&26 \\
5474 \ion{C}{II}&  6 & 0&1:& 0&14: &  \multicolumn{1}{r}{\nodata}
  &\multicolumn{2}{c}{\nodata} & \multicolumn{2}{c}{\nodata}  \\
5540 \ion{C}{II}&  2 & 0&04:&  0&06:&  2& 0&13 & 0&14\\
5656 \ion{C}{II}&  24&0&48& 0&52 & 12 &0&77 & 0&79 \\
5696 \ion{C}{III}($\log$ flux)&40 & $-$14&07 & $-$12&40 & 21& $-$13&70$^{\rm c}$& $-$12&72$^{\rm c}$ \\
5806 \ion{C}{IV} & 8& 0&22& 0&19 & 7 &0&25& 0&23\\  
5826 \ion{C}{III} &  14& 0&50& 0&47&  14 &0&79 & 0&75 \\
5876 \ion{He}{I} & 20 & 0&70& 0&63 & 29 & 1&39 & 1&31 \\
5880 \ion{C}{II} & 18 & 0&80& 0&66 & 24 & 1&00 & 0&89 \\ 
5894 \ion{C}{II} & 10 &0&40& 0&30 & 8 & 0&50 & 0&45 \\
6097  \ion{C}{II}& 13& 0&45& 0&37 & 14 &0&88 & 0&74 \\
6151  \ion{C}{II}& 15 & 0&58& 0&47& 14 & 0&90 & 0&70 \\
6206 \ion{C}{III}& 4 &0&10 & 0&07 & 3 & 0&19 & 0&14 \\
6255  \ion{C}{II}& 7 & 0&18& 0&11 & 4 &0&28 & 0&21 \\
6462  \ion{C}{II}& 17& 0&62& 0&37 & 18 & 0&92 & 0&64 \\
6512     & 4 & 0&17& 0&09 & 3 &  0&1:&   0&07: \\
6578  \ion{C}{II}& 67& 3&20 & 1&65 & 54 &2&29 & 1&46 \\
6678 \ion{He}{I} & 37& 1&06& 0&52 & 20 &1&48 & 0&92 \\
6746  \ion{C}{II}& 54& 1&70 & 0&84 & 6 & 0&37 & 0&21 \\
6789 \ion{C}{III}&  57& 2&53& 1&19 & 26& 1&85 & 1&03 \\
7035 \ion{C}{II} &  16& 0&73& 0&30 & 12 &1&90 & 0&97 \\
7065 \ion{He}{I} &  50& 2&16& 0&81 & 21 &2&99 & 1&52 \\
7116 \ion{C}{II} &  35& 1&48& 0&55 & 20 &2&59 & 1&28 \\
7235\tabnotemark{b} \ion{C}{II} & 280& 14&2& 5&27& 129& 19&0 & 9&15 \\
7282 \ion{He}{I} & 13&  0&78& 0&29 & 6 & 0&52 & 0&27 \\
7580 \ion{C}{III} &10 &  0&45& 0&14 & \multicolumn{1}{r}{\nodata} & \multicolumn{2}{c}{\nodata}
 & \multicolumn{2}{c}{\nodata} \\
\bottomrule
\multicolumn{7}{l}{$^{\rm a}$ Equivalent width in \AA.}\\
\multicolumn{7}{l}{$^{\rm b}$ Uncertainties are discussed in the text.} \\
\multicolumn{7}{l}{$^{\rm c}$ Value for September 2000. See discussion in \S 5.1.}\\
\end{tabular}
\end{table*}

\setlength{\tabnotewidth}{0.8\textwidth}
\tablecols{4} 
\begin{table*}[!t]\centering
\caption{Stellar line parameters for spectral classification}
\begin{tabular}{lcccc}
\toprule
\noalign{\smallskip}
                      & HuDo\,1   & HuBi\,1 & [WC\,10]\tabnotemark{a} \\
\midrule
\noalign{\smallskip}
$\log$ F(\ion{C}{3} 5696)\tabnotemark{b}& $-$12.40 & $-$12.77 & \nodata \\
FWHM(\ion{C}{3} 5696) (\AA)           & 14       & 7 & 3--6 \\
$\log$ F(\ion{C}{4} 5808)/F(\ion{C}{3} 5696) & $-0.72\pm0.10$ & $-0.64\pm0.10$ & $-$1.2 to $-$0.7\\
$\log$ F(\ion{C}{4} 5808)/F(\ion{C}{2} 4267) & $-0.69\pm0.13$ & $-0.73\pm0.13$ & $-$1.5 to $-$0.2 \\
$\log$ F(\ion{He}{2} 4686)/F(\ion{He}{1} 5876)& $-0.50\pm0.20$ & $-0.82\pm0.20$ & $\leq$ $-$0.8 \\
\noalign{\smallskip}
\bottomrule
\tabnotetext{a}{ Classification criteria by Crowther et al. (1998).} 
\tabnotetext{b}{ Dereddened fluxes in erg cm$^{-2}$ s$^{-1}$.} 
\end{tabular}
\vspace{0.4cm}
\end{table*}

\setlength{\tabnotewidth}{0.7\textwidth}
\tablecols{6}
\begin{table*}[tb]\centering
\caption{Nebular and stellar parameters for all\linebreak
 studied [WC~10] PNe}
\begin{tabular}{lrrrrr}
\toprule
\noalign{\smallskip}
 $\log$(X/H)+12 & HuDo\,1 & HuBi\tabnotemark{a}& M4--18\tabnotemark{b}
 & CPD--56\tabnotemark{c} & He2--113\tabnotemark{c} \\
 \midrule
 \noalign{\smallskip}
  He         & $>$10.8 & $>$11.04 & \nodata & \nodata & \nodata \\ 
  O          & 8.43    & 8.57   & 8.62 & 8.68 & 8.68 \\
  N          & 7.75    & 7.61   & 7.60 & 7.91 & 7.82 \\
  C          &         &        & 9.08 & 9.80 & 9.70  \\
  \hline 
  \noalign{\smallskip}
 N$_{\rm e}$ (cm$^{-3}$) & 3300   & 800    & 5011 & 60,000 & 60,000 \\
 $T_{\rm e}$ (K)& \nodata & 9400   & 8200 & 9300  & 8400 \\
 diameter ($''$)  & $\sim$1.7& 18  & 1.85    & 1.7$\times$2.1& 1.4$\times$1.1\\
 c(H$\beta$)      & 2.05    & 1.22 & 0.81 & 1.00 & 1.47 \\
 $\log$ F(H$\beta$)$^{(d)}$  & $-$12.75 & $-$12.59 & $-$11.43& $-$11.03 & $-$10.21 \\
 distance (kpc)          & \nodata & \nodata & 6.8 &   1.35  & 1.2 \\
 $V_*$(\tabnotemark{d})   & 14.4     & 15.8 & 12.5 & 9.5 & 8.9 \\
 $T_*$  (kK)    & \nodata & 35     & 31  & 30    & 30   \\
 $\log$ \, $L_*/L_\odot$ & \nodata & (3.7)   &3.72 & 3.7 & 3.72\\
 $\log$ ($\dot M/M_\odot$ yr$^{-1}$) & \nodata  &  $-$5.70   & $-$6.05 & $-$5.4 & $-$6.1 \\
 v$_\infty$ (km s$^{-1}$)& \nodata &  360  & 160 & 225 & 160\\
\noalign{\smallskip}
   \bottomrule 
\tabnotetext{a}{Stellar wind data for HuBi\,1 are from Leuenhagen \& Hamann (1998).}
\tabnotetext{b}{Data for M4-18 are from De Marco \& Crowther (1999).}
\tabnotetext{c}{Data for CPD-56$^0$8032 and He2--113 are from  De Marco et~al.\@ (1997).}
\tabnotetext{d}{F(H$\beta$) and stellar $V$ magnitudes have been corrected for reddening.}
\end{tabular}
\end{table*}


The stellar line fluxes, relative to \ion{C}{3}~$\lambda$5696, were dereddened
assuming the Seaton (1979) reddening law and a logarithmic reddening correction at
H$\beta$ as derived from the nebular lines.  Once dereddened we proceeded to
classify the stars following the criteria proposed by Crowther et~al.\@  (1998) (see Table~5).
We have not used Acker \& Neiner (2003) scheme because, as mentioned in the
Introduction, their suggested criteria show discontinuities and the primary ones
are not well defined.

Unfortunately, the primary criteria proposed by Crowther et~al.  (and also Acker
\& Neiner's), fall mainly on the \ion{C}{4} 5801,5812 lines, which are extremely
faint for very late [WC] stars.  In addition, for our spectral resolution these
lines are almost blended with (or at least lie in the wing of) the strong
\ion{C}{3} 5826 and show P\,Cygni profiles (see Fig.~4), difficulting their
measurements.  All this causes the main uncertainty in our classification of
HuDo\,1 and HuBi\,1 central stars.  However, within uncertainties, both objects
can be classified as [WC\,10] stars, although the FWHM(\ion{C}{3} 5696) for
HuDo\,1 is too large.  Considering this primary criterion and taking into
account the (very uncertain) \ion{He}{2}/\ion{He}{1} ratio, the central star in
HuDo\,1 could be nearer the [WC\,9] spectral type, but all the other primary
criteria point to the [WC\,10] type.  In general, the equivalent widths and the
FWHM of lines in HuDo\,1 are larger than in other [WC\,10] stars, probably
indicating a more massive wind with a larger expansion velocity or larger
turbulence.  In this sense, HuDo\,1 is a peculiar [WC\,10] object, in a similar
sense to the peculiar [WC\,9] SwSt\,1 which has narrower stellar lines than other
[WC\,9] objects (Crowther et~al.\@  1998).


\subsection{ Is the Central Star of HuBi\,1 Variable?}

In Table 4 we present for HuBi\,1 a dereddened $\log$ F(\ion{C}{3} 5696) = $-$12.72,
from the B\&Ch data obtained in September 2000.  From the literature, we found
that, from observations performed in July 1996, Crowther et~al.\@  (1998) reported
a dereddened $\log$~F(\ion{C}{3} 5696) = $-$11.85, while Acker \& Neiner (2003)
reported a value of $-$12.24, from observations performed between 1994 and 1995.
Our observations of June 2002, show a fainter star, with $\log$~F(\ion{C}{3}
5696) = $-$13.50. However the stellar line ratios seem not to change
significantly.  Of course much of the flux variations could be caused by the
differences in the slit widths or observing conditions, but it would be worthy
to perfom a long-term photometric analysis of HuBi\,1, in order to verify the
possibility of stellar variations.  It is known that the [WC\,10] star
CPD-56$^0$8032 varies with $\Delta$V$\sim$1 mag (Pollacco et~al.\@  1992) and small
variations ($\Delta$V$\sim$~0.03 mag) have been reported for M4--18 (Handler
1996).


\section{Discussion and results}

From stellar spectroscopic data and applying recent criteria for WR spectral
classification, we have found a spectral type [WC\,10] (peculiar) for HuDo\,1
and we confirmed a [WC\,10] type for HuBi\,1.  This locates these objects among
the coolest [WC] stars.

Interestingly, both PNe, but in particular HuBi\,1, show relatively low nebular density
for such late [WC]-type objects.  Several authors (e.g., Acker, Gorny, \&
Cuisinier1996; Gorny
\& Tylenda 2000; Pe\~na et~al.\@  2001) have shown that nebulae around [WC] stars
follow a density-spectral type relation, with the electron densities
decreasing from nebulae around late [WC] stars (young objects showing N$_{\rm e}
\geq 10^4$ cm$^{-3}$) to nebulae around early [WC] stars (evolved objects).
Nevertheless there are some exceptions to this rule.  The case of HuBi\,1 was
already pointed out by Pe\~na et~al.\@  (2001).  This nebula and K2--16 ([WC\,11]
central star) are the only [WC-late] ones presenting densities lower than 10$^3$ cm$^{-3}$.
It has been suggested that these kind of objects could represent born-again
planetary nebulae or, alternatively, low-mass slowly-evolving stars.  Gorny \&
Tylenda (2000) and Pe\~na et~al.\@  (2001) have argued in favor of the latter
suggestion.  The case of HuDo\,1 (also M4--18, see below) is only marginally out
of the density-[WC] type relation.

In Table 6 we present a compendium of the known parameters for the up-to-now 5
well-studied PNe around [WC\,10] central stars.  For HuDo\,1, we have estimated
the dereddened apparent visual magnitude, V$_*$, from the continuum flux at 5500
\AA.  The value for HuBi\,1 is from Hu \& Bibo (1990) after dereddening.  An
analysis of Table~6 gives the following results:

(1). All the PNe have O/H abundance ratio between 8.4 and 8.7, typical of disk
PNe (see Kingsburgh \& Barlow 1994).  The N/O ratio is between 0.1 and 0.2,
which is slightly low for typical PNe but similar to the solar value.  That is,
none of these nebulae seems to be N-rich (curiously, none Peimbert Type~I PN is
found in this sample, although there are several in the [WC-early] types), and
the only three for which C has been measured seem very C-rich.  Regarding He, we
found that HuBi\,1 is the only nebula ionized by a cool [WC] star, showing He$^+$
recombination lines.  This indicates that HuBi\,1 is very probably a He-rich
nebula and its central star should be slightly hotter than the other objects,
for which most of the nebular helium is in neutral form and its abundance cannot
be measured.

(2).  The apparent magnitudes for Hudo\,1 and HuBi\,1 central stars are much
fainter than the values of the other [WC\,10] stars (HuBi\,1 has the apparent
faintest star).  Assuming that all of them have a similar luminosity would imply
that HuBi\,1 and HuDo\,1 should be at longer distances.  However if we suppose a
distance of 6.8~kpc (the same as for M4--18) for HuBi\,1, the nebular diameter
would be 0.6~pc, which seems too large for a nebula with an apparently young
central star.  Therefore, at least for HuBi\,1 we should conclude that its star
is intrinsically fainter and, consequently, of lower mass.  This represents an
evidence that the masses of WR central stars even with very similar effective
temperatures, span a large range of post-AGB masses.  The same conclusion was
derived by Pe\~na et~al.\@  (1998) from the analysis of nebulae with [WC-early]
stars.

(3). Due to its large diameter, higher excitation and low density HuBi\,1 seems
the more evolved object, with probably the largest dynamical age.  It is the
only object for which a ``born-again'' scenario is plausible but, as suggested
by Pe\~na et~al.\@  (2001), most probably it is  a low-mass
slowly-evolving post-AGB object.  There are some indications pointing that
HuBi\,1 star could be variable.  This particular object certainly deserves a
close follow-up.

\acknowledgments
Professional and technical support from the members of the staff at Observatorio
Astron\'omico Nacional, M.  Richer, F.  Montalvo, and G. Garc{\'\i}a, is deeply acknowledged.  This
work received partial financial support from CONACYT-M\'exico and DGAPA-UNAM
(grant IN114601).


\end{document}